\documentclass[prb,superscriptaddress,twocolumn]{revtex4-2}
\usepackage[utf8]{inputenc}     
\usepackage[dvips]{graphicx}      
\usepackage{amsmath,bm}
\usepackage{xcolor}
\usepackage{multirow}
\usepackage{graphicx}
\usepackage{multirow}
\usepackage{bigstrut}

\usepackage[paperwidth=8.5in, paperheight=11in]{geometry}
\usepackage{color, soul}

\RequirePackage[hyperfigures=true, hyperfootnotes=true, hyperindex=true,breaklinks=true]{hyperref}
\definecolor{link_green}{rgb}{0.0,0.7,0.0}
\definecolor{link_blue_dark}{rgb}{0.0,0.0,0.7}
\definecolor{link_blue}{rgb}{0.0,0.0,1}
\definecolor{link_red}{rgb}{0.7,0.0,0.0}
\definecolor{link_red_dark}{rgb}{0.6,0.0,0.0}
\definecolor{link_red_verydark}{rgb}{0.3,0.0,0.0}
\hypersetup{hyperfigures=true,hyperfootnotes=true,hyperindex=true,pagebackref,breaklinks=true,citecolor=link_green,filecolor=link_blue,linkcolor=link_blue_dark, urlcolor=link_blue,colorlinks,bookmarks,bookmarksopen,bookmarksnumbered, pdfauthor="Olivier Fruchart",pdfstartview=FitH} \RequirePackage[all]{hypcap}

\usepackage{siunitx}%
\RequirePackage{xspace}
\RequirePackage{upgreek}%
\RequirePackage{journames}

  \RequirePackage[normalem]{ulem} 
  \RequirePackage{soul}
  \definecolor{olivier_comment}{rgb}{0.6,0.6,0.9}
  \definecolor{olivier_add}{rgb}{0.8,0.2,0.2}
  \definecolor{olivier_remove}{rgb}{0.5,0.5,0.5}

\RequirePackage[mathcal]{euscript}

\def\eg{{\it e.g.}\/\xspace}%
\def\ie{{\it i.e.}\/\xspace}%
\newcommand{\vect}[1]{\mathbf{#1}}
\def\muZero{\mu_{\mathrm 0}}%
\def\Ks{K_{1}}
\def\Kf{K_{2}}
\def\Ms{M_{\mathrm s}}
\def\vectm{\vect m}
\def\HOE{H_{\mathrm{\OE}}}
\def\vectHOE{\vect{H}_\mathrm{\OE}}
\newcommand{\OErsted}{\OE rsted\/\xspace}%
\newcommand{\lex}{\ell_\mathrm{ex}}
\newcommand{\lOE}{\ell_\mathrm{OE}}
\newcommand{\muB}{\mu_\mathrm{B}}
\newcommand{\DeltaT}{\Delta_\mathrm{T}}
\newcommand{\JW}{J_\mathrm{W}}
\newcommand{\Rext}{R_\mathrm{e}}
\newcommand{\Rint}{R_\mathrm{i}}

\renewcommand{\partial}{\mathrm{d}}

\begin{document}

\title{ Theoretical study of current-induced domain wall motion in magnetic nanotubes with azimuthal domains, including Œrsted field and spin-transfer torques}

\author{J\'{e}r\^{o}me Hurst}
\email{jerome.hurst@cea.fr}
\affiliation{Univ. Grenoble Alpes, CNRS, CEA, Grenoble INP, Spintec, 38000 Grenoble, France}

\author{Arnaud De Riz}
\affiliation{Univ. Grenoble Alpes, CNRS, CEA, Grenoble INP, Spintec, 38000 Grenoble, France}

\author{Michal Sta{\v{n}}o}
\affiliation{CEITEC, Brno University of Technology, Brno,  61200 Brno
Czech Republic}

\author{Jean-Christophe Toussaint}
\affiliation{Univ. Grenoble Alpes, CNRS, Institut NEEL, 38000 Grenoble, France}

\author{Olivier Fruchart}
\affiliation{Univ. Grenoble Alpes, CNRS, CEA, Grenoble INP, Spintec, 38000 Grenoble, France}

\author{Daria Gusakova}
\email{daria.gusakova@cea.fr}
\affiliation{Univ. Grenoble Alpes, CNRS, CEA, Grenoble INP, Spintec, 38000 Grenoble, France}

\date{\today}

\begin{abstract}
We report a theoretical overview of the magnetic domain wall behavior under an electric current in infinitely-long nanotubes with azimuthal magnetization, combining the $1$D analytic model and micromagnetic simulations. We highlight effects that, besides spin-transfer torques already largely understood in flat strips, arise specifically in the tubular geometry: the \OErsted field and curvature-induced magnetic anisotropy resulting both from exchange and material growth. Depending on both the geometry of the tube and the strength of the azimuthal anisotropy, Bloch or Néel walls arise at rest, resulting in two regimes of motion largely dominated by either spin-transfer torques or the \OErsted field. We determine the Walker breakdown current in all cases, and highlight the most suitable parameters to achieve high domain wall speed.

\end{abstract}

\maketitle

\section{Introduction}

Magnetic nanowires and to a lesser extent nanotubes have been synthesized for three decades, and mostly investigated magnetically as large assemblies\citep{bib-FER1999a,bib-SOU2014,bib-VAZ2015,bib-FRU2018d}. The number of investigations at the scale of single objects has been sharply increasing in recent years, through sensitive analytical techniques\citep{bib-WEB2012,bib-GRO2016}, magnetic microscopies\citep{bib-HEN2001,bib-BIZ2013,bib-KIM2011b,bib-STR2014}  or electric transport\citep{bib-DOU1997,bib-EBE2000,bib-RUeF2012,bib-MOH2016,bib-GIO2020}. The focus has been largely devoted to magnetic domain walls~(DWs), motivated by the prediction of existence of a new type of DW in soft-magnetic nanowires and nanotubes with head-to-head domains, characterized by the absence of Walker breakdown and thus high DW speed\citep{bib-THI2006,bib-YAN2012}, and a new dynamic regime with emission of spin waves\citep{bib-YAN2011b}. The picture of DWs at rest\citep{bib-BIZ2013,bib-FRU2014} and their quasistatic motion\citep{bib-PAL2015,bib-FRU2016c,bib-MOH2016,bib-BRA2018} has been confirmed experimentally in wires, however, reports on their mobility and precessional dynamics are only emerging\citep{bib-FRU2019b}.

While theory predicts that domains are longitudinally-magnetized in nanowires and nanotubes made of a soft-magnetic material, in recent years several experimental reports pointed at the possible existence of magnetic domains with fully or partly azimuthal magnetization, at least at the outer periphery of the object, both wires\citep{bib-BRA2017,bib-RUI2018}  and tubes\citep{bib-STR2014,bib-FRU2018g}. The dipolar field minimization in the finite-length\citep{bib-WYS2017,bib-SUN2014,bib-SKO2000b} or diameter modulated wires\citep{bib-CHA2012,bib-SAL2013b,bib-IVA2016b} may promote such stated. However, the unambiguous observation of azimuthal magnetization in very long nanotubes can only be explained by the existence of a microscopic contribution to the magnetic anisotropy, favoring the azimuthal and not longitudinal magnetization direction, \eg, through magnetostriction.

DWs emerge between two such domains with opposite circulation, which do not move under the application of a uniform magnetic field\citep{bib-RUI2018}, as the Zeeman energy of both domains is the same. It is thus expected that only an electric current may set such DWs into motion. Interestingly, besides the conventional spin-transfer torques, a significant \OErsted field directly coupled to azimuthal magnetization through Zeeman energy may exist in a tubular geometry. Its amplitude may reach several tens of mT for nanotube diameter of few tens of nanometers and for current densities comparable to those required for the spin-transfer-torque-induced DW motion, \eg, of the order of $\SI{e+12}{\ampere\per\meter\squared}$. Curvature-induced anisotropy together with \OErsted field effects suggest that according to the DW type their dynamical features may be very different from the case of thin flat strips, now well established theoretically and experimentally\citep{bib-THO2007,bib-THI2008}. While the effect of the \OErsted field has been reported for the DW nucleation at the ends of axially-magnetized wires\citep{bib-OTA2015,bib-FER2020} and drafted for the DW motion in these structures\citep{bib-AUR2013}, there exists no overview of the \OErsted field effect on the DW motion in nanotubes with azimuthal magnetization.

It is our purpose here, to draw the general picture of current-driven DW motion in the case of azimuthal domains in the tubular geometry, ahead of experimental reports. We consider nanotubes and not nanowires, and with outer diameter below~$\SI{100}{\nano\meter}$, for the sake of simplicity. This way, we expected to extract unambiguously the physics of DWs in relation with size and curvature, and with the strength of azimuthal anisotropy. We combine analytical modeling and micromagnetic simulations to draw the panorama of the statics of DWs, and their dynamics under applied current including spin-transfer torque effects and the Zeeman effect of the \OErsted field.

\newpage

\section{Theoretical framework}

\subsection{General micromagnetic framework}
\label{Subsection:micromagnetic_framework}

Domains and DWs in a ferromagnetic material are usually described within the framework of the micromagnetic theory \citep{bib-BRO1963b}, which is based on a continuous description of the unit magnetization vector ${\bf{m}}$ and of all other quantities. In the presence of a spin-polarized electron flow through magnetic texture, the time evolution of ${\bf{m}}$ is governed by the Landau-Lifshitz-Gilbert (LLG) equation\citep{bib-LAN1935}, generalized with the so-called adiabatic and nonadiabatic spin torques\citep{bib-THI2005}:
\begin{align}
\frac{\partial {\vectm}}{\partial t} &= \frac{\gamma_0}{\muZero\Ms} {\vectm} \times \frac{\delta \mathcal{E}\left[{\vectm}\right]}{\delta {\vectm}} + \alpha  {\vectm} \times \frac{\partial {\vectm}}{\partial t}  \nonumber \\
&-\left[{\bf{U}} \cdot {\bf{\nabla}} \right] {\vectm} +  \beta {\vectm} \times \left(\left[{\bf{U}} \cdot {\bf{\nabla}} \right] {\vectm}\right),
\label{Eq:LLG}
\end{align}
with $\gamma_0=\muZero|\gamma|$ the gyromagnetic ratio, $\alpha$ the phenomenological Gilbert damping coefficient, and $\mathcal{E}\left[\vectm\right]$ the functional of volume density of energy of the system, which may include exchange, magnetostatic, anisotropy and \OErsted field contributions. The spin-transfer torque contributions are proportional to the electric current ${\bf{J}}$ via ${\bf{U}} = -\muB P {\bf{J}} / (e \Ms)$ with $P$ the spin-polarization ratio, $\muB$ the Bohr magneton, $e$ the (positive) elementary charge, $\beta$ the non-adiabatic coefficient.

In most cases the functional for the volume density of energy has a non-trivial dependence on spatial coordinates, and requires a fully numerical treatment. Nevertheless, under some conditions it may be simplified and thus the behavior of the magnetic texture may be predicted analytically. In the next section we described aspects specific to the case of the tubular geometry with a thin shell.

In this paper we combine analytical results with numerical solutions of Eq.\eqref{Eq:LLG}, obtained using the home-built finite element freeware $\textit{feeLLGood}$\citep{bib-ALO2014,bib-STU2015,bib-FEE}. We consider the material parameters of Permalloy Ni$_{80}$Fe$_{20}$ as a prototypical soft magnetic material: exchange stiffness $A = \SI{13}{\pico\joule\per\meter}$, $\muZero\Ms = \SI{1}{\tesla}$ the spontaneous induction, and $P$=0.7 the spin polarization of conduction electrons. We use $\alpha=1$ to reach equilibrium configurations, and $\alpha$=0.02, $\beta$=0.04 to describe the dynamics of DWs under current.

\subsection{Micromagnetism in the tubular geometry}
\label{Subsection:Analytical_description}

\begin{figure}[t]
{
\includegraphics[width=8cm]{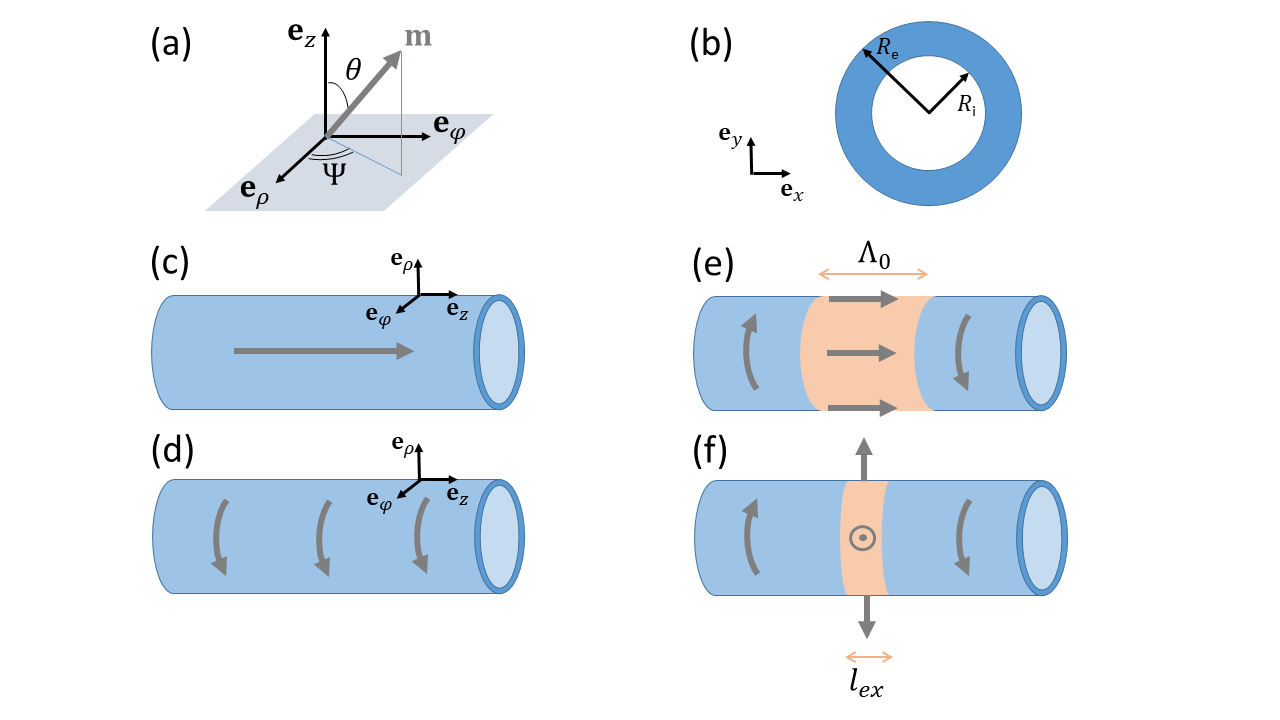}}
\caption{(a)~Sketch of the basis of cylindrical coordinates describing positions in space, and the two angles defining magnetization in a spherical basis. (b) Sketch of the nanotubes cross section. Illustration of four magnetic textures: (c)~axial single domain; (d)~azimuthal single domain, (e)~Néel-like DW; (f)~Bloch-like DW. Here $l_{ex}$ and $\Lambda_0$ stands respectively for the dipolar exchange length and the effective anisotropy exchange length and are defined later in the text.}
\label{fig:Sketch_system}
\end{figure}

In the following we make use of the cylindrical basis (${\bf{e}}_{\rho}$, ${\bf{e}}_\phi$, ${\bf{e}}_z$) for referencing locations in space. We described the unit magnetization vector ${\vectm}$ with spherical coordinates, fully determined by two angles: polar $\theta=\theta\left( \rho,\phi,z \right)$, and azimuthal $\Psi=\Psi \left( \rho,\phi,z \right)$ as follows:
\begin{align}
{\bf{m}} &=
\begin{pmatrix}
m_{\rho}  \\
m_{\phi}\\
m_{z}
\end{pmatrix}
=
\begin{pmatrix}
 \sin\theta \cos \Psi \\
 \sin\theta\sin\Psi\\
 \cos\theta
\end{pmatrix},
\label{Eq: Definiton magnetization cylindrical coordinates}
\end{align}
These notations are illustrated on Fig.\ref{fig:Sketch_system}(a), and are common for the tubular geometry\citep{bib-THI2006,bib-LAN2010,bib-FRU2018d}.

An azimuthal component of magnetization corresponds to $\Psi\neq0$. It gives rise to a curvature-induced contribution to the exchange energy inversely proportional to the square of the nanotube's radius\citep{bib-LAN2010}. This favors alignment of magnetization along the nanotube axis, to avoid this energy. Therefore, the experimental observation of azimuthal domains in long nanotubes, for which magnetostatic energy associated with the nanotube's ends can be disregarded to favor azimuthal magnetization such as in curling states\citep{bib-SUN2014,bib-WYS2017}, hints at the existence of an extra energy term favoring the azimuthal direction. Here we will describe it phenomenologically as its microscopic origin is not proven at present, either magnetostriction combined with curvature-induced anisotropic strain, or inter-grain interface anisotropy combined with anisotropic grain shape. Azimuthal anisotropy may be taken into account either as an easy-axis anisotropy contribution along the azimuthal direction $-K_\phi m_\phi^2$ with $K_\phi>0$, or as a hard-axis anisotropy contribution along the nanotube axis~$K_z m_z^2$ with $K_z>0$. From Eq.\ref{Eq: Definiton magnetization cylindrical coordinates} it appears that the two descriptions are equivalent only for $\Psi=\pi/2$, with $K_z=K_\phi$, \ie, in the absence of radial component. Experimentally, the anisotropy field associated with the azimuthal anisotropy has been found not to exceed a few tens of mT so far\citep{bib-FRU2018g}, which is negligible against the cost of radial magnetization, analogous to a perpendicularly-magnetized film in rolled-strip picture for a nanotube, see Fig.\ref{fig:Sketch_system}). However, this difference must show up in critical situations, \eg at the transition from Néel to Bloch DWs upon increasing the nanotube thickness~(analogous to the thickness of a thin film). In the present work we make the choice of a hard-axis anisotropy contribution along the nanotube axis. This slightly favors radial magnetization against the choice of an easy-axis anisotropy contribution along the azimuthal direction, for a given value of~$K$, so that the transition from Néel to Bloch DWs is expected to occur at a slightly lower nanotube thickness. This must be kept in mind, as we will see in the following that the type of DW occurring has a crucial impact on their dynamics. Besides, for the sake of completeness, in the present work we consider also values of anisotropy much larger than those reported so far, so that the difference between the two choices becomes even more significant.

In the following, we include the phenomenological hard-axis anisotropy term in the so-called thin-shell analytical model detailed in Ref.\citenum{bib-LAN2010}. The principle of this model is to evaluate the magnetostatic energy by neglecting the contributions of  magnetic volume charges and considering only the surface penalty for magnetization pointing along the normal to the cylindrical surface, analogous to a rolled thin film. Therefore, the model is best suited to describe thin nanotubes. In this case the functional for the volume density of energy reads:
\begin{align}
E\left[ {\vectm}  \right] 
&=\int \mathrm{d}V \mathcal{E}_0\left[{\bf{m}}\right] \nonumber \\
&= \int \mathrm{d}V \bigg{[} A \left( \bm{\nabla} \vectm\right)^2 + 
 \frac{\mu_0 \Ms^2}{2}  m_\rho^2  +  K m_z^2 \bigg{]} .
\label{Eq: MicroMag General Energy Functional}
\end{align}

To describe the DW dynamics we work in the frame of a $1$D model, boiling down to a $z$-dependance only of all quantities. This takes advantage of the cylindrical symmetry, assuming an azimuthal invariance, and also no variation along the radius, which is reasonable in the case of thin nanotubes. Therefore, $ \theta\left( \rho,z,\phi \right)$ and $\Psi\left( \rho,z,\phi \right)$ become $\theta(z)$ and $\Psi(z)$. The energy functional Eq.\eqref{Eq: MicroMag General Energy Functional} becomes
\begin{align}
E\left[ {\bf{m}}  \right] 
&= S A \int \mathrm{d}z\bigg{[} \left(\partial_z \theta \right)^2+
\sin^2 \theta  \left(\partial_z \Psi \right)^2 \nonumber \\
&+ \frac{\sin^2 \theta  }{\lambda^2} + \frac{\sin^2 \theta \cos^2 \Psi}{l_{ex}^2}+ \frac{\cos^2\theta}{W^2}
\bigg{]} \nonumber \\
&= S  \int dz~\mathcal{E}_0\left[\theta(z),\Psi(z)\right],
\label{Eq:Energy functional 1D model for ground state}
\end{align}
with $W = \sqrt{A/K}$, $\lex = \sqrt{2A/\muZero \Ms^2}$ the dipolar exchange length, $S = \pi \left( \Rext^2-\Rint^2 \right)$ the surface of the nanotube section and $\lambda = \sqrt{ \left( \Rext^2 - \Rint^2 \right)/(2\ln(\Rext/\Rint))}$. Here, $\Rext$ and $\Rint$ denote, respectively, the external and the internal radius of the nanotube. The two first terms in the integral \eqref{Eq:Energy functional 1D model for ground state} correspond to the exchange energy commonly found in $1$D DW dynamics models such as for flat strips, whereas the third term is the curvature induced exchange energy, specific to the tubular geometry. The parameter $1/\lambda$ may be seen as the curvature parameter, which value is large for small radius and approaches zero when the radius becomes infinite. Also, $\lambda$ is similar to the radius for thin-shell tubes~($\Rext\approx\Rint$). The fourth term in Eq.(\ref{Eq:Energy functional 1D model for ground state}) corresponds to the demagnetization energy and the last term is the contribution of the uniaxial anisotropy.

For the sake of future discussion on the residual quantitative differences between this $1$D model and full micromagnetic simulations, let us summarize the key approximations of the $1$D model: absence of radial dependance of magnetization, and magnetostatic energy taken into account as a local ultrathin-film-like term.

\section{Domains and domain walls at rest}
\label{Sec:Magnetic stationary states}


Following the standard procedure of the energy functional minimization [Appendix \ref{Appendix:Euler_equation}] we obtain a set of two differential equations describing equilibrium magnetic distributions:
\begin{align}
&\theta''=  \sin \theta\cos \theta \left[ \Psi'^2 - \frac{1}{\Lambda_0^2} + \frac{\cos^2\Psi}{\lex^2} \right],
\label{Eq:set of differential equation theta}  \\
&\Psi'' \sin \theta + 2 \Psi' \cos \theta    = -\frac{\sin \theta \sin \left(2 \Psi \right)}{2\lex^2},
\label{Eq:set of differential equation psi}
\end{align}
with $\Lambda_0^2 = A / \left( K-\Ks \right)$ and $\Ks = A / \lambda^2$. $\Lambda_0$ is the effective anisotropy exchange length for azimuthal magnetization. It is not straightforward to find analytical magnetization profiles satisfying both Eqs. \eqref{Eq:set of differential equation theta}-\eqref{Eq:set of differential equation psi}.  Nevertheless, it is possible to find some particular solutions.

\paragraph{Azimuthal versus axial domains.}

Two trivial stable solutions are $\left\{\theta = 0 ~;~ \Psi =cst \right\}$ and $\left\{\theta = \pi/2 ~;~ \Psi = \pi/2 \right\}$. These correspond to axial~[longitudinal, Fig. \ref{fig:Sketch_system}(b)] and azimuthal~[Fig. \ref{fig:Sketch_system}(c)] domains, respectively.

\begin{figure}[t]
\centering
\includegraphics[scale=0.28]{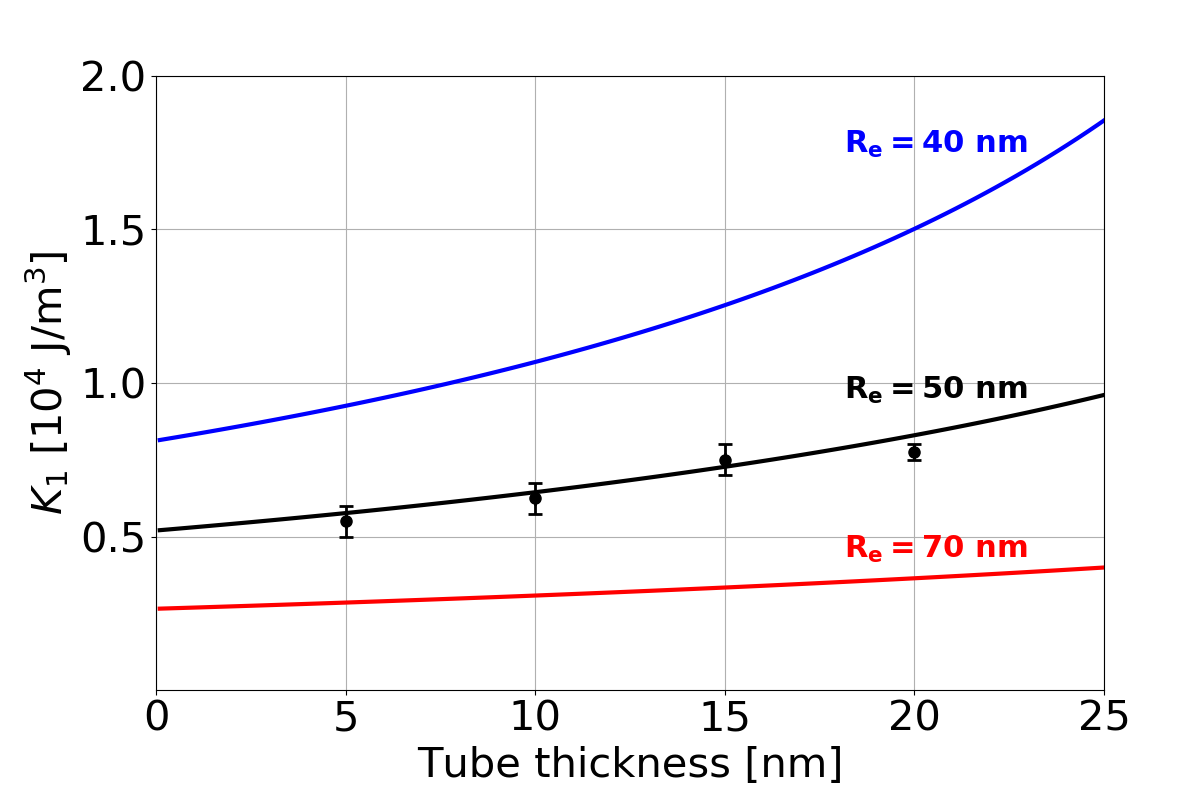}
\caption{Threshold values of the hard-axis anisotropy strength $\Ks$ as a function of the nanotube thickness for three different external radius. The markers with the errorbars correspond to an estimation of $\Ks$ obtained with micromagnetic simulations. }
\label{fig:ks}
\end{figure}

The total energy $E_\mathrm{tot}$ of the axial single domain state is $VK$, against $VA/\lambda^2$ for the azimuthal single domain state. Here, $V = LS$ is the volume of the nanotube, $S$ being the area of its cross-section and $L$ its length. The ground state is therefore axial magnetization for $K<\Ks$, and azimuthal magnetization for $K>\Ks$, which is well known\citep{bib-FRU2018d}. The underlying physics is already clear from the examination of the third and fifth terms on the right-hand side of Eq.\ref{Eq:Energy functional 1D model for ground state}: noticing that $\sin^2\theta=1-\cos^2\theta$, curvature-induced exchange and the azimuthal anisotropy of microscopic origin play competing roles. They are exactly balanced for $W=\lambda$, and thus for $K=\Ks$. Fig.\ref{fig:ks} shows $\Ks$ versus the nanotube thickness and for different nanotube diameters.

To set an order of magnitude, $\Ks$ is of the order of mT for realistic material parameters and radius: $\Ks \approx\SI{8}{\milli\tesla}$ for $\Rext = \SI{50}{nm}$ and $\Rint = \SI{40}{nm}$, dropping \citep{bib-STR2014,bib-ZIM2018}.

\paragraph{Néel-like and Bloch-like domain walls.}

Here we focus on domain walls in infinitely-long tubes, separating two domains with azimuthal magnetization of opposite directions, which from the above we expect for $K>\Ks$. Note that the case of domains with axial magnetization, such as expected for $K<\Ks$, has been already well described\citep{bib-FOR2002,bib-HER2002a,bib-THI2006,bib-FRU2015b}: domain walls may be of either transverse type or azimuthal type~(also called curling or vortex wall). The former is promoted for small diameter and thick shell, while the latter is promoted for large diameter and thin shell.

On a theoretical basis, profiles of domain walls shall be found for a uniform value $\Psi=\pi/2$ or $\theta=\pi/2$ along the entire tube. On this basis, we shall consider two cases to reduce the partial differential equation describing magnetization~[Appendix \ref{Appendix:1D_model_profile}].

For $\Psi = \pi/2$, we find the following magnetization profile:
\begin{align}
m_{\rho}=0,~m_{\phi} = \pm \tanh\left( \frac{z}{\Lambda_0}\right),~m_z=\cosh^{-1}\left( \frac{z}{\Lambda_0}\right),
\label{Eq:Azimuthal_Neel_wall_formula}
\end{align}
This is a DW of length $\Lambda_0$, for which magnetization remains parallel to the tube surface at every location~[Fig.\ref{fig:Sketch_system}(e)]. By unrolling the surface of the cylinder, mapping it to a flat strip aligned along $z$ and width $2\pi R$ with periodic boundary conditions, we notice that this corresponds to a $\ang{180}$ Néel DW, so that we will refer to it as a Néel DW in the following.

For $\theta = \pi/2$, we find the following magnetization profile:
\begin{align}
m_{\rho} =  \cosh^{-1}\left( \frac{z}{\lex}\right),~~m_{\phi}  = \pm \tanh\left( \frac{z}{\lex}\right),~~m_{z}=0,
\label{Eq:azimuthal_Bloch_wall_formula}
\end{align}
This is a DW of length $\lex$, for which magnetization is perpendicular to the surface of the nanotube at the center of the DW~[Fig.\ref{fig:Sketch_system}(f)]. By unrolling again the surface of the cylinder, mapping it to a flat strip, this corresponds to a $\ang{180}$ Bloch DW, so that we will refer to it as a Bloch DW in the following.

Within the $1$D model, the thickness of Néel and Bloch walls is given, respectively, by the effective anisotropy exchange length $\Lambda_0$ and the dipolar exchange length $l_{ex}$. This is a direct consequence of the approximated dipolar energy used in Eq. \eqref{Eq: MicroMag General Energy Functional} together with the $1$D model assumption which allow Néel walls (resp. Bloch walls) to be, independent of the strength of the dipolar field (resp. the effective anisotropy field).

\begin{figure}[t]
\centering
\includegraphics[width=8cm]{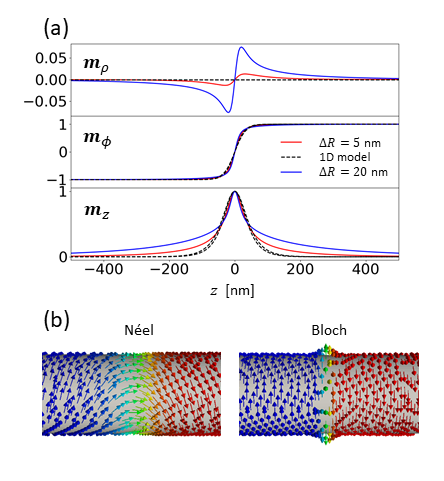}\caption{
 (a)  N\'{e}el DW magnetization profile along $z$ direction. Black dashed lines correspond to the equation \eqref{Eq:Azimuthal_Neel_wall_formula}. Red and blue lines correspond to micromagnetic profiles averaged over $\rho$ and $\phi$ for two nanotube shell thicknesses: $\Delta R$=5nm and $\Delta R$=20nm. The following parameters have been used: $\alpha = 1$, $K = 2 \times 10^{4}~ \rm{J/m^3}$, $R_e = 50 ~\rm{nm}$, $L=$1500nm. (b) Micromagnetic distribution of N\'{e}el and Bloch DW.}
  \label{fig:Ground_state_magnetization_anaytic_vs_simulations}
\end{figure}

\begin{figure}[t]
\centering
\includegraphics[width=8cm]{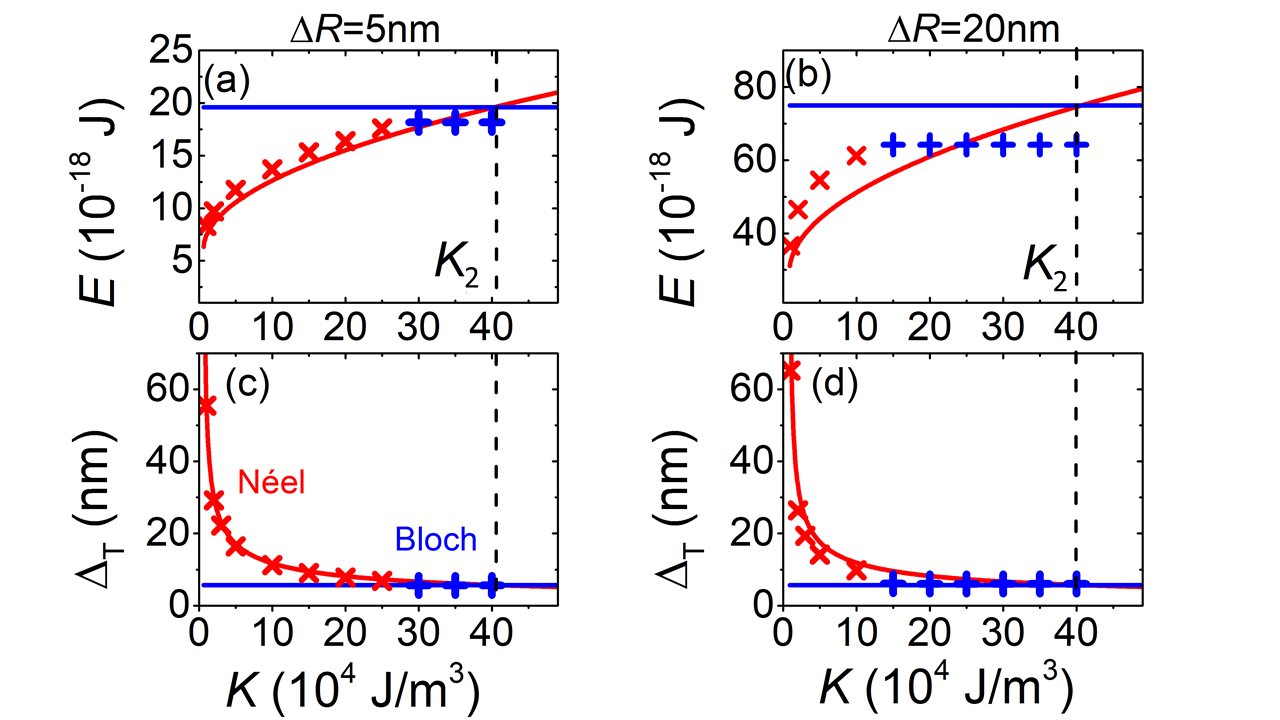}
  \caption{(a-b)  Total energy and (c-d) DW length as a function of the hard-axis anisotropy parameter $K$. Solid lines correspond to the $1$D model and crosses to micromagnetic simulations. The red~(resp. blue) color highlights the case of Néel~(resp. Bloch) DW being of lower energy in the micromagnetic simulations, thus the ground state under static conditions. For all graphs the following parameters have been used: $\alpha$=1, $\Rext=\SI{50}{\nano\meter}$, $L=\SI{700}{\nano\meter}$, while $\Delta R=\SI{5}{\nano\meter}$ and $\Delta R=\SI{20}{\nano\meter}$ in the left and right columns, respectively. Vertical dashed line highlight $\Kf$ values expected from the $1$D model. }
  \label{fig:Energy_analytic_vs_smulations_ground_state}
\end{figure}

In order to characterize both DWs versus the anisotropy coefficient~$K$, we calculate their total energy using Eq.\eqref{Eq:Energy functional 1D model for ground state}, and their length using the so-called Thiele definition\citep{bib-THI1973,bib-THI2006}:
\begin{align}
\DeltaT = \frac{2S}{\int \mathrm{d}z \left[\left( \partial_z m_{\rho}\right)^2+\left( \partial_z m_{\phi}\right)^2 +\left( \partial_z m_{z}\right)^2\right]}.
\label{Eq:Thiele_length}
\end{align}
The Thiele definition is of particular relevance for the discussion of DW dynamics, as will be considered in the next section, and of practical interest to compare the $1$D model with fully $3$D distributions resulting from micromagnetic simulations. For the $1$D model considered here, \eqref{Eq:Thiele_length} simplifies to $\DeltaT = \Lambda_0$ for the Néel DW and $\DeltaT = \lex$ for the Bloch DW.

Fig.\ref{fig:Ground_state_magnetization_anaytic_vs_simulations} shows an example of DW profile, comparing the $1$D model with 3D micromagnetic simulations. The $1$D model appears fairly faithful, especially for thin-shell tubes. Fig.\ref{fig:Energy_analytic_vs_smulations_ground_state} allows a global comparison, displaying the total energy and the Thiele length for both types of DWs as a function of the hard-axis anisotropy coefficient~$K$. This figure highlights the existence of a cross-over value for anisotropy, for which both DWs have the same energy, and incidentally the same length. This is a direct consequence of the tradeoff between anisotropy (and to a lesser extent azimuthal exchange) energy disfavoring longitudinal magnetization in the core of a Néel Wall and out-of-plane-like magnetostatic energy dominating in the core of a Bloch DW. This explains that the $1$D model predicts the cross-over for $\Kf = A (1/\lambda^2+1/\lex^2)$. For $K<\Kf$ the Néel DW is the ground state. Its width $\Lambda_0$ results from the competition between exchange and magnetostatic energy of the DW. It scales with $1/(K-\Ks)$, diverging at $K=\Ks$, reflecting the softening of effective anisotropy and the continuous transition from  a Néel DW to an axial mono domain. For $K>\Kf$, the Bloch DW is the ground state. Its energy and length are independent of the strength of axial anisotropy, as there are no magnetic moments pointing along the axial direction. According to the $1$D model, the nanotube thickness has an impact on the DW energy, however not changing significantly the cross-over between the Néel and Bloch DWs. Indeed, the threshold  values of the anisotropy $\Ks$ and $\Kf$ depend on the nanotube thickness, however their variation remains moderate: $\Ks =\SI{0.57E4}{\joule\per\cubic\meter}$ and  $\Kf =\SI{40.37E4}{\joule\per\cubic\meter}$ for $\SI{5}{\nano\meter}$ thickness and $\Ks =\SI{0.81E4}{\joule\per\cubic\meter}$ and $\Kf =\SI{40.62E4}{\joule\per\cubic\meter}$ for $\SI{20}{\nano\meter}$ thickness.

Micromagnetic simulations agree reasonably well with the $1$D model for DW energies and lengths for thin-shell nanotubes ($\Delta R <\SI{5}{\nano\meter}$), while the discrepancy is more pronounced for thicker shell nanotubes. This has a sizeable impact on the threshold anisotropy value $\Kf$, which the simulations find lower than the $1$D model predicts. For instance in micromagnetics, Bloch DWs are already the ground state for $K = \SI{15E4}{\joule\per\cubic\meter}$ and $\SI{20}{\nano\meter}$ thickness, while the $1$D model predict a value around $\SI{40E4}{\joule\per\cubic\meter}$. This difference is less pronounced for thinner shells, but still significant.

We attribute this discrepancy to the demagnetization energy, which is imperfectly taken into account in the $1$D model, as a local term and focusing on surface magnetic charges, not volume magnetic charges. In the case of nanotubes, volume charges can be approximated by neglecting, respectively, the variation of $m_\rho$ and $m_\phi$ along the radial and azimuthal direction as follows:
\begin{equation}
\rho_m=-\Ms\bm{\nabla} \cdot \mathbf{m}\approx -\Ms(m_{\rho}/\rho  + \partial m_{z} / \partial z).
\label{Eq:volumeCharges}
\end{equation}
In Eq.(\ref{Eq:volumeCharges}), the first term results from the curvature. It matters for Bloch DWs, however depends only weakly on the thickness of the shell. Indeed, it simply reflects a shift of surface charges from the inner surface to the volume, to account for the unequal inner and outer surface and preserve the total amount of charges of the Bloch DW, so that in the end it is already reasonably taken into account in the $1$D model. The second term shows up specifically in the Néel DW, giving rise to an internal demagnetizing field, and is not considered in the $1$D model for the Néel. The physics revealed by the micromagnetic simulations is therefore analogous to the transition from Bloch to Néel DWs in thin films upon lowering their thickness\citep{bib-NEE1955}, resulting from a tradeoff between vertical versus planar demagnetizing fields. Therefore, we expect that for nanotubes of large diameter and therefore negligible curvature effects, the transition from Néel to Bloch or even cross-tie DWs\cite{bib-STR2014} is found for a value of thickness to that in thin films\citep{bib-HUB1998b}. Thus, while the trends for DWs in nanotubes with azimuthal magnetization can be understood with an analogy to a rolled thin film and the $1$D model, Fig.\ref{fig:Ground_state_magnetization_anaytic_vs_simulations} reveals an effect specific to nanotubes:  in micromagnetic simulations, the radial component of magnetization $m_{\rho}$ in a Néel DW is non-zero, an effect growing with the shell thickness. The reason is the existence of volume charges, which were already known to induce a radial tilt of magnetization for Bloch-point DWs in nanowires\citep{bib-THI2006} and vortex DWs in nanotubes\citep{bib-YAN2012,bib-LAN2010} with head-to-head magnetization. For these DWs the tilt is monopolar, while for a Néel DW the tilt is bipolar, as it is associated with a dipolar distribution of charges. An alternative view is that, mathematically, $\rho_m$ is decreased as $\partial m_{z} / \partial z $ and  $m_{\rho}$ have opposite signs.

\section{1d model of domain wall motion}

In this section we investigate the dynamics of Néel and Bloch DWs by means of a $1$D model, the variational formulation of the LLG equation, and the collective coordinate method. We consider the dynamics driven by a homogeneous charge electric current flowing along the axial direction of the nanotube ${\bf{J}} = J\;{\mathbf{e}}_z$. The current couples to magnetization through two effects. First, it induces an \OErsted field in the entire nanotube, and second, it gives rise to spin-transfer torques in the DW.\OErsted field have been considered theoretically, for example, to assist magnetization reversal in cylindrical core-shell nanostructures\cite{bib-OTA2015}. We take into account both effects.

\subsection{\OErsted field}

The \OErsted field within the nanotube shell ($\Rint \leq \rho <  \Rext$) reads
\begin{align}
\vectHOE\left(\rho\right) ={\bf{e}}_{\phi}J \rho \left[1- (\Rint/\rho)^2\right]/2,
\label{Eq:Oersted field in tube}
\end{align}
which modifies the energy density functional (Eq.\ref{Eq:Energy functional 1D model for ground state}) as follows:
\begin{align}
&\mathcal{E}\left[ \theta(z), \Psi(z)  \right] =\mathcal{E}_0 - \frac{2 A\sin \theta \sin  \Psi }{\lOE^2}.
\label{Eq:energy functional simplified + oersted}
\end{align}
Here we introduced the characteristic length $\lOE = \sqrt{2A/(\mu_0 \Ms \HOE)}$, with $\HOE$ defined as the averaged value of the \OErsted field inside the nanotube shell
\begin{align}
\HOE= J\overline{R}= \frac{J\left(\Rext-\Rint\right)\left( \Rext+2\Rint\right)}{3 \left(\Rext+\Rint\right)}.
\label{Eq:def Oersted field averaged on tube}
\end{align}
The quantity $\lOE$ could be called the \OErsted exchange length, balancing exchange and \OErsted-Zeeman energy, in analogy to the dipolar exchange length~$\lex$. To set an idea, the \OErsted field at the external nanotube surface is  6-$\SI{20}{\milli\tesla}$  for a current density $J = \SI{1E12}{\ampere\per\meter\squared}$ in a nanotube of radius $\SI{50}{\nano\meter}$ and shell thickness $\Delta R= 5-\SI{20}{\nano\meter}$. While no experiments of current-induced DW motion have been reported in nanotubes to date, the latter parameters for radius and current density are similar to the case of DW motion in cylindrical nanowires reported experimentally\cite{bib-FRU2019b}.

\subsection{Equation of motion}

In order to quantify the DW motion we use so-called collective coordinate method \citep{bib-LAN2010,bib-THI2005}, in which the whole magnetic texture is described with a small number of macroscopic variables. This simplifies the resolution of the dynamics of magnetization using the variational formulation of Eqs.\eqref{Eq:LLG} and \eqref{Eq:energy functional simplified + oersted}, detailed in Appendix \ref{Appendix:Equation_of_motion}. The core of the collective variables method is to find a suitable ansatz for the DW profile, to get an exact expression for the action and the dissipation function of the system\citep{bib-SLO1972,bib-THI2004}. This allows one to construct a tractable set of differential equations for each time-dependent collective variable. Assuming that the DW remain rigid during motion, we introduce the coordinate of the center of the DW $z_0(t)$, its length $\Lambda(t)$ and its radial tilt $\chi(t)$. For our purpose we set the following ansatz for the DW profile:
\begin{align}
m_{\rho} &= \sin\chi(t) \cosh^{-1}\left[ \frac{z - z_0(t)}{\Lambda (t)}\right], \nonumber \\
 m_{\phi} &= \tanh\left[ \frac{z - z_0(t)}{\Lambda (t)}\right], \nonumber \\
m_{z}  &= \cos\chi(t) \cosh^{-1}\left[ \frac{z - z_0(t)}{\Lambda (t)}\right].
\label{Eq:Anzatz_dynamics_azimuthal_neel_wall}
\end{align}
The ansatz \eqref{Eq:Anzatz_dynamics_azimuthal_neel_wall} allows to describe Néel and Bloch DWs, depending on the value of $\chi$. At $t=0$, the collective variables are initialized to their values at rest:   $z_0 = 0$, $\Lambda = \Lambda_0$, $\chi = 0$ for the Néel DW, and $z_0 = 0$, $\Lambda = \lex$, $\chi = \pi/2$ for the Bloch DW.\\

After some algebra~(see Appendix \ref{Appendix:Equation_of_motion}), we obtain the following equations of motion for each collective variable:
\color{black}
\begin{align}
 \frac{\dot{\Lambda}}{\Lambda} &=
 \frac{12\gamma_0A}{\alpha \pi^2 \mu_0\Ms}
       \left(\frac{1}{\Lambda^2}  -   \frac{\cos^2 \chi}{\Lambda_{0}^2}  - \frac{ \sin^2\chi}{\lex^2} \right),
       \label{Eq:equation of motion collective variables model lambda}
         \\
\dot{z_0} &=
\frac{\dot{\chi}\Lambda}{\alpha} - \frac{ \gamma_0 \Lambda }{\alpha}\HOE + \frac{\beta }{\alpha}U,
\label{Eq:equation of motion collective variables model zo}
 \\
\dot{\chi} &=
 \frac{1}{\tau (1+\alpha^2)} \left[ \frac{2\HOE}{\Ms} - \alpha\sin \left( 2\chi \right) \left( 1 - \frac{\lex^2}{\Lambda_{0}^2} \right)\right]  \nonumber \\
 & ~~ +\frac{U}{\Lambda(\alpha+1/\alpha)} \left( 1-\frac{\beta}{\alpha}\right),
\label{Eq:equation of motion collective variables model chi}
\end{align}
with $\tau=2/(\gamma_0 M_s)$ and ${\bf{U}} = -\muB P {\bf{J}} / (e \Ms)$.

These equations are analogous to $1$D models for the combined field and current-driven motion of DWs with axial domains, in nanostrips\citep{bib-THI2004,bib-TRE2008} or for vortex DWs in thin-shell nanotubes\citep{bib-OTA2012,bib-LAN2010}. Indeed, if we permute the magnetization and current-induced magnetic field from the azimuthal to the axial direction and vice versa, we recover the system behaviour described previously. However, new terms related to the hard-axis anisotropy contribution may play a significant role, in particular $(1-\lex^2/\Lambda_{0}^2)$. Besides, in contrast to the case of external uniform longitudinal magnetic field case reported previously\citep{bib-LAN2010}, the strength of the \OErsted field is linked to the charge current flowing through the nanotube, so that the corresponding driving force cannot be considered separately from that of spin-transfer torques. 

\begin{figure}[t]
\centering
\includegraphics[scale=0.35]{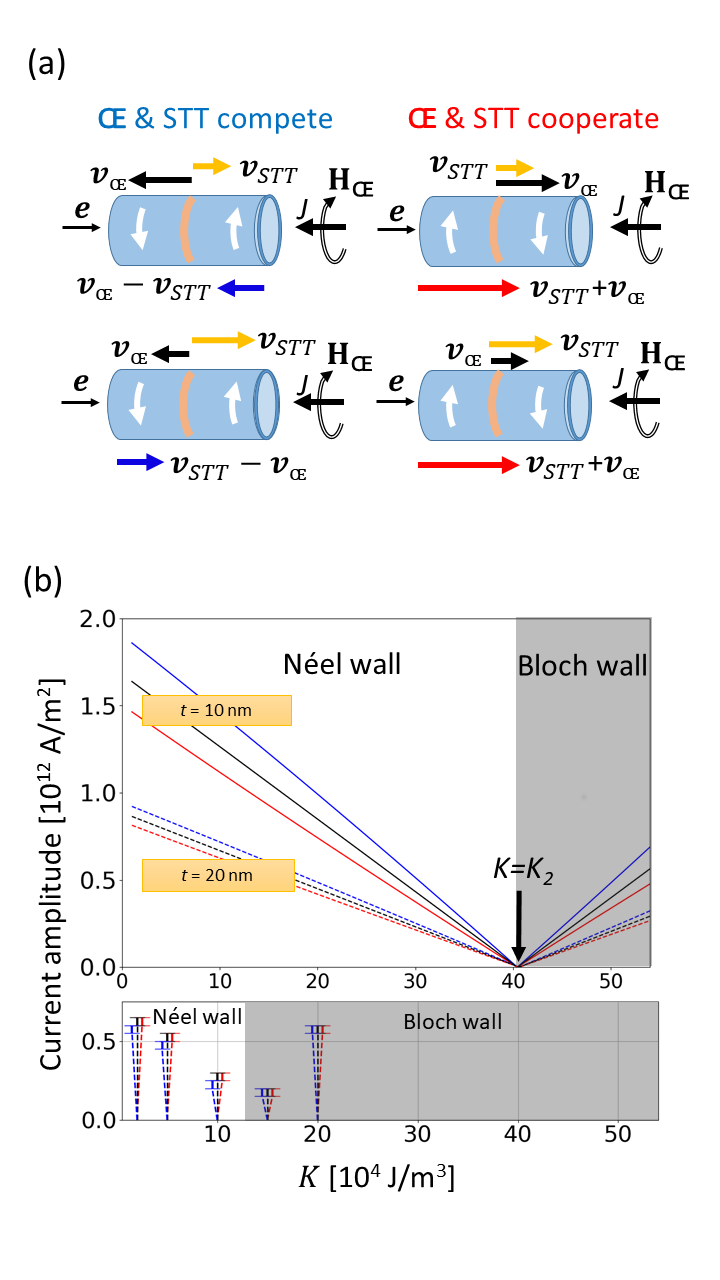}
  \caption{ (a) Schematics for the cases of competing and cooperating \OErsted and spin-transfer torques, depending on the relative sign of domains circulation and applied current.  For each configurations, we sketch two possible situations: an \OErsted field dominated dynamics (upper plot) and a spin-transfer-torque-dominated dynamics (lower plot). (b) Upper plot: absolute value of $\JW$ given by Eq.\eqref{Eq: Critical Oersted field} versus the magnitude of anisotropy $K$ for two nanotube thicknesses: $\SI{10}{\nano\meter}$ (solid lines) and $\SI{20}{\nano\meter}$ (dashed lines). Lower plot: absolute value of $\JW$ versus $K$ obtained in micromagnetic simulations for $\SI{20}{\nano\meter}$ shell thickness. In both cases, black color stands for the omission of spin-transfer torques, while red~(resp. blue) stand for cooperating~(resp. competing) \OErsted and spin-transfer torques. The external radius value is $\Rext = \SI{50}{\nano\meter}$.}
  \label{fig:shema_1D_model_walker_field}
\end{figure}

Similar to the case of strips, a low current enables a steady-state solution with $\dot{\chi}=0$, while above a cross-over value of currents a so-called Walker regime sets in, with $\dot{\chi} \neq 0 $.  However, in contrast to flat strip and depending on the circulation of the azimuthal domains with respect to the direction of current, two distinct situations occur. In the first case the \OErsted field and the spin-transfer torques cooperate, meaning that they tend to move the DW in the same direction. In the second case they compete. Indeed, spin-transfer always promote motion along the flow of electrons, while the effect of the \OErsted field depends on the relative circulation of the two domains~[Fig. \ref{fig:shema_1D_model_walker_field}(a)]. We will see that this competition has consequences both in the steady state and Walker regimes, for both Néel and Bloch~DWs, although quantitative differences show up depending on the DW type.

\subsection{Walker regime}

The critical value of the current $J_{\textrm{w}}$ separating the two dynamical regimes, steady-state and Walker, reads
\begin{align}
\JW &= \frac{\alpha  }{\mu_0 M_s \overline{R}} \chi_0 \left( \frac{\mu_0 M_s^2}{2} - \left( K-\Ks\right) \right) \nonumber \\
&\hspace{2.5cm} \left[1- C_{\pm}\frac{\tau \mu_B P}{2 \Lambda_{\textrm{w}} e \overline{R}} \left(\alpha-\beta\right) \right]^{-1},
\label{Eq: Critical Oersted field}
\end{align}
with $\Lambda_{\textrm{w}} = \left[ 1/(2\Lambda_0^2) + 1/(2\lex^2) \right]^{-1/2}$, $\chi_0=+1$ for an inial Néel DW and $\chi_0=-1$ for an initial Bloch DW. $C_{\pm}=+1$~(resp. $-1$) for cooperating~(resp. opposing) \OErsted field and spin-transfer torques. The two effects are clearly separated in Eq.(\ref{Eq: Critical Oersted field}), in the first and second brackets: that of the effective azimuthal anisotropy, and that of the spin-transfer torque acting through the polarization of conduction electrons. This is illustrated on Fig.\ref{fig:shema_1D_model_walker_field}(b), and discussed below.

The critical current varies linearly with the anisotropy $K$, decreasing in the region with Néel DW as ground state, and increasing in the region with a Bloch DWs as ground state. This reflects the internal restoring force of the DW against the excitations, driving it away from equilibrium during the dynamics. The cross-over from the Néel to the Bloch ground states highlights a soft mode, obviously associated with no restoring force and thus Walker regime for any applied current. In this very specific case, the DW moves along the $z$ axis at constant linear velocity, with a core precessing in the $\rho-z$ plane at constant angular velocity without changing its length. This case is analogous to the purely precessional motion of transverse-vortex DWs in cylindrical nanowires\citep{bib-FOR2002b,bib-NIE2002,bib-THI2006,bib-YAN2010}. In the general case the DW speed is oscillating, however for very large currents $J \gg J_{\textrm{w}}$ it is possible to derive an analytic formula for the time-averaged DW speed  from Eqs. \eqref{Eq:equation of motion collective variables model lambda}-\eqref{Eq:equation of motion collective variables model chi}:
\begin{align}
\langle \vert \dot{z_0} \vert \rangle_t = \frac{\alpha \gamma_0  \overline{\Lambda} \HOE}{1 + \alpha^2} + C_{\pm} \frac{U \left( 1 +\beta \alpha \right)}{1 + \alpha^2}\;,
\label{Eq:Velocity_domain_wall_large_current}
\end{align}
where $\overline{\Lambda}=\Lambda_{\textrm{w}}$ corresponds to the time averaged DW length. In this regime and for realistic values of $\alpha$, the \OErsted field dependent term is negligible compared to the SST dependent term so that the direction of motion is imposed by the electron flow. The physics is analogous to strips, with a combination of motion induced by current and magnetic field, except that those cannot be fixed independently in our system since they both originate from the current density.

Micromagnetic simulations confirm this picture qualitatively~(Fig.\ref{fig:shema_1D_model_walker_field}(b)). Quantitatively, the $\Kf$ values for the transition of Bloch to Néel DWs at rest are lower than in the $1$D model, with a discrepancy increasing with the shell thickness. As discussed in section \ref{Sec:Magnetic stationary states}, we attribute this differences to magnetic volume charges neglected in the $1$D model, further promoting Bloch DWs such as in thin flat films.

\subsection{Steady-state regime}

In the low current regime $J < J_{\textrm{w}}$, the out-of-plane angle $\chi$ converges in time to a constant value. During this process, the DW progressivly accelerates and contracts, until reaching the steady-state regime with a constant velocity and length. For very low currents ($J \ll J_{\textrm{w}}$) one derives:
\begin{align}
&\Lambda \simeq   \Lambda_{0}~~\textrm{and}~~
|\dot{z_0}| \simeq  \frac{ \gamma_0 \Lambda_0 \HOE}{\alpha}  + C_{\pm}  \frac{\beta U}{\alpha} ~~\textrm{(N\'{e}el)},
\label{Eq:Velocity_domain_wall_small_Orsted_field_Neel} \\
&\Lambda \simeq   l_{ex}~~\textrm{and}~~
|\dot{z_0}| \simeq  \frac{ \gamma_0 \lex \HOE}{\alpha}  + C_{\pm}  \frac{\beta U}{\alpha} ~~\textrm{(Bloch)},
\label{Eq:Velocity_domain_wall_small_Orsted_field_Bloch}
\end{align}
where Eq.\eqref{Eq:Velocity_domain_wall_small_Orsted_field_Neel} stands for Néel DWs and Eq.\eqref{Eq:Velocity_domain_wall_small_Orsted_field_Bloch} stands for Bloch DWs. Once again, Eqs.\eqref{Eq:Velocity_domain_wall_small_Orsted_field_Neel}-\eqref{Eq:Velocity_domain_wall_small_Orsted_field_Bloch} 
are similar to the case of DW motion in strips, except that the sources of anisotropy and magnetic field are different. Both types of DW are in a steady-state regime, with two contributions to the velocity: \OErsted-field-induced and spin-transfer-torque-induced. The first term is proportional to the DW width, thus its strength very sensitively on the DW type, playing a more or less important role in comparison with the second term. It means that in some particular cases one may observe either \OErsted-field dominated or spin-transfer-torque-dominated dynamics. For this reason, below we examine in more detail the dynamics  of Néel and Bloch DWs, combining $1$D model equations of motion \eqref{Eq:equation of motion collective variables model lambda}-\eqref{Eq:equation of motion collective variables model chi} solved with a standard Runge-Kutta solver (RKK4), and micromagnetic simulations.

\section{Dynamics of Néel walls}

\begin{figure}[t]
\centering
\includegraphics[scale=0.26]{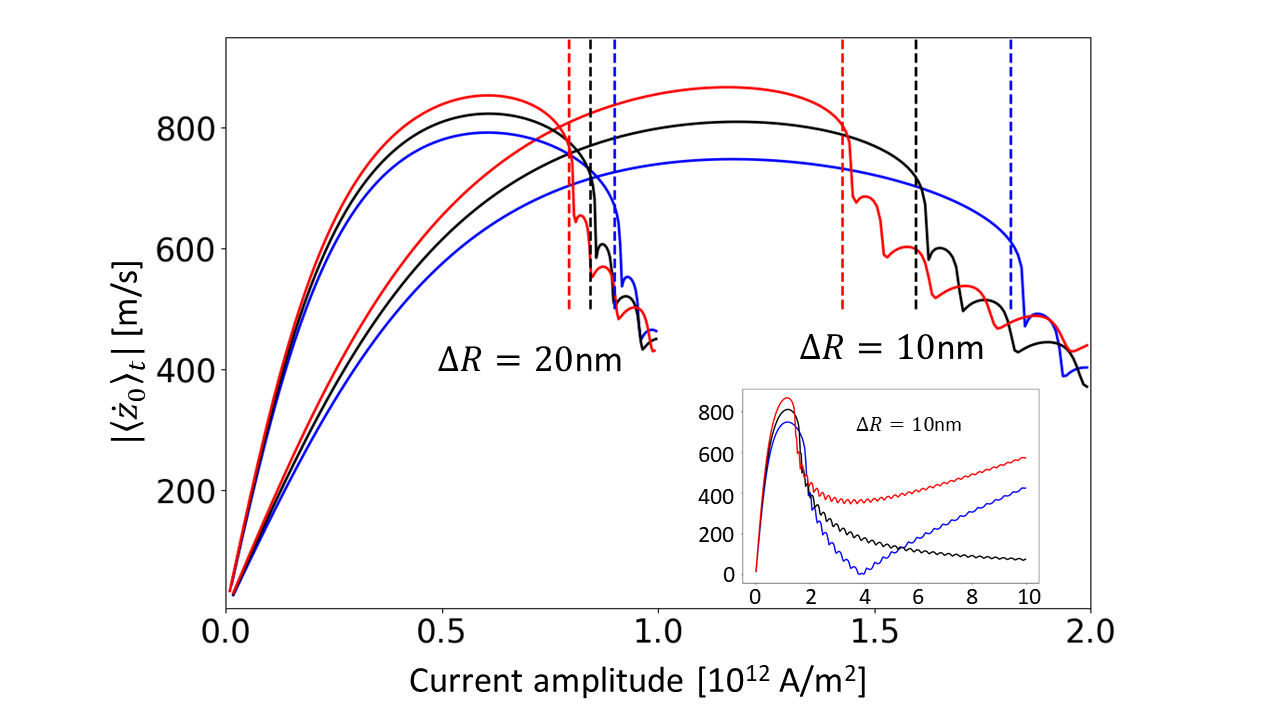}
  \caption{Time-averaged DW speed amplitude according to Eqs.\eqref{Eq:equation of motion collective variables model lambda}-\eqref{Eq:equation of motion collective variables model chi} for a N\'{e}el DW, versus the density of current. We considered two thicknesses of shell~($\Delta R=\SI{10}{\nano\meter}$ and $\Delta R=\SI{20}{\nano\meter}$) and three different situations schematized in Fig.\ref{fig:shema_1D_model_walker_field}(a): black color for absence of spin-transfer torques, red color for cooperating \OErsted and spin-transfer torques, and blue color for competing \OErsted and spin-transfer torques. The vertical dashed lines highlight the value of critical current $\JW$ defined by Eq.\eqref{Eq: Critical Oersted field}. The following parameters have been used: $K = \SI{2E4}{\joule\per\meter\squared}$ and $\Rext = \SI{50}{\nano\meter}$. The inset plot shows the time-averaged DW speed amplitude on a larger current range with a shell of $10$ nm of thickness. }
  \label{fig: Average_velocity_vs_alpha_1D_model_azimuthal_neel_wall}
\end{figure}

\begin{figure}[t]
  \centering
\includegraphics[scale=0.28]{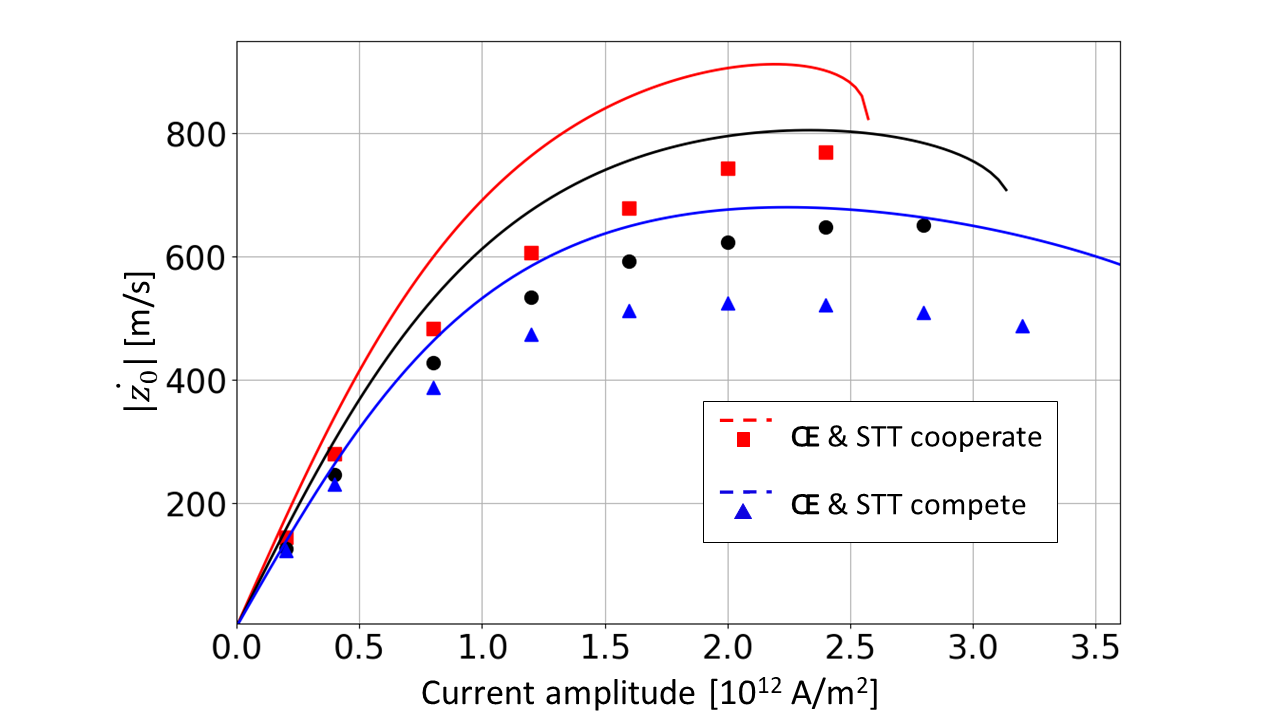}
   \caption{Comparison of the steady velocity amplitude of a Néel wall, between the $1$D model~(solid lines) and micromagnetic simulations~(symbols), versus the density of current. Three situations schematized in Fig.\ref{fig:shema_1D_model_walker_field}(a) were considered: black color for absence of spin-transfer torque, red color for cooperating \OErsted and spin-transfer torques, and blue color for competing \OErsted and spin-transfer torques. The following parameters have been used: $K = \SI{2E4}{\joule\per\meter\squared}$, $\Rext = \SI{50}{\nano\meter}$ and $\Rint = \SI{45}{\nano\meter}$.}
 \label{fig: DW_speed_vs_current_and_STT}
\end{figure}

In the steady-state regime, Néel DWs may demonstrate relatively high velocities . Indeed, their dynamics in this regime is largely dominated by the \OErsted field contribution, related to a relatively large DW width~[Fig. \ref{fig:Energy_analytic_vs_smulations_ground_state}(c,d)] for moderate anisotropy $\sim\SI{1E4}{\joule\per\cubic\meter}$, such as typical for the experimental situation reported so far\citep{bib-FRU2018g}. Under these conditions the dipolar restoring force preventing the transformation of a Néel DW into a Bloch DW is large, and the maximal speed that a DW may reach is around $\SI{800}{\meter\per\second}$ for the material and geometrical parameters considered here, and almost independent of the nanotube thickness. However, the mobility is larger for thicker-shell nanotubes, as in this case the average \OErsted field induced in the magnetic material is higher for a given density of current. In turn, this translates in a lower Walker current~$\JW$. The effect of spin-transfer torques is moderated in those conditions~(Fig.\ref{fig: Average_velocity_vs_alpha_1D_model_azimuthal_neel_wall}), for all nanotube shell thicknesses studied here. Qualitatively similar behavior of steady velocities have been observed in micromagnetic simulations on Fig.\ref{fig: DW_speed_vs_current_and_STT} although the values of velocity are somewhat lower. 
To visualize the impact of competing and cooperating \OErsted and spin-transfer torque, in Figs. \ref{fig: Average_velocity_vs_alpha_1D_model_azimuthal_neel_wall} and \ref{fig: DW_speed_vs_current_and_STT}  we have plotted the absolute value of the DW velocity, the direction of DW motion being towards positive $z$ ( resp. negative $z$ ) for cooperating ( resp. competing) configuration as shown in Fig. \ref{fig:shema_1D_model_walker_field}(a) for \OErsted dominated dynamics. 

\begin{figure}[t]
  \centering
\includegraphics[scale=0.28]{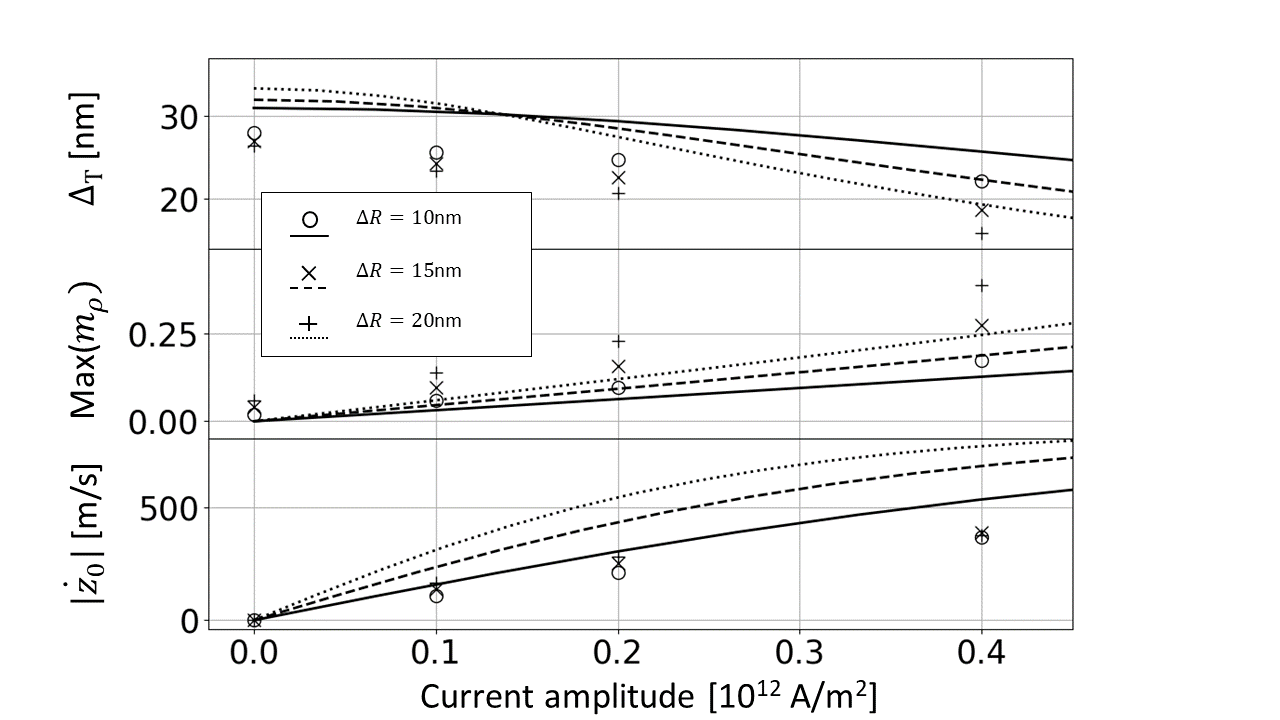}
  \caption{Domain wall length (upper panel), maximum out-of- plane magnetization (middle panel) and DW speed amplitude (lower panel) as a function of the current amplitude for three nanotube thicknesses and purely \OErsted-field-induced dynamics. Lines correspond to the $1$D model, and symbols to micromagnetic simulations. The simulations were performed with following parameters: $\Rext = \SI{50}{\nano\meter}$, $P=0$  and $K = \SI{2E4}{\joule\per\cubic\meter}$.}
 \label{fig: DW_speed_vs_current_and_thickness}
\end{figure}

To understand the origin of the quantitative difference between the $1$D model and micromagnetic simulations, we examine the DW length $\DeltaT$, maximum of the radial magnetization component $m_{\rho}$, and steady-state DW velocity~(Fig.~\ref{fig: DW_speed_vs_current_and_thickness}). For a better visibility we kept only those curves corresponding to purely \OErsted-field-induced dynamics, thus without spin-transfer torques. With increasing current the radial component $m_{\rho}$ increases and the DW width $\DeltaT$ decreases, the latter being analogous to thin films. As in a field-dominated regime the DW speed is proportional to the DW width~[Eq.\eqref{Eq:Velocity_domain_wall_small_Orsted_field_Neel}], this translates in a convex variation of DW speed with current. Nevertheless, the DW length is systematically overestimated by the $1$D model. We believe that it is due to the omission of the demagnetizing field, thus failing to grasp the narrow core of Néel walls\citep{bib-HUB1998b}, predominant for the Thiele definition of the DW width. This is most likely responsible for the DW speed overestimation in the 1D model for the \OErsted field dominated dynamics. Note that in spite of the high velocities that can be reached in this field-dominated dynamics, the direction of motion of DWs is dictated by the sign of circulation of azimuthal magnetization in the two domains~[Fig. \ref{fig:shema_1D_model_walker_field}(a)]. This means that two consecutive walls in a nanotube move along opposite directions. This is probably not desirable for most concepts of devices, implying the absence of propagation of information, and the possibility for DW annihilation. Finally, it must be stressed that the critical current $\JW$ and the maximum reachable speed decrease with increasing and large anisotropy.

Above the critical current $\JW$, depicted with the vertical dashed lines in Fig.\ref{fig: Average_velocity_vs_alpha_1D_model_azimuthal_neel_wall}, the average speed drops sharply and shows an oscillatory decreasing behavior typical for Walker regime. For very high current densities, unrealistic experimentally, the DW  speed rise again with a mobility proportional to $U$, see Eq. \eqref{Eq:Velocity_domain_wall_large_current}, and is nearly driven by spin-transfer torque.

\section{Dynamics of Bloch walls}

\begin{figure}[t]
\centering
\includegraphics[scale=0.28]{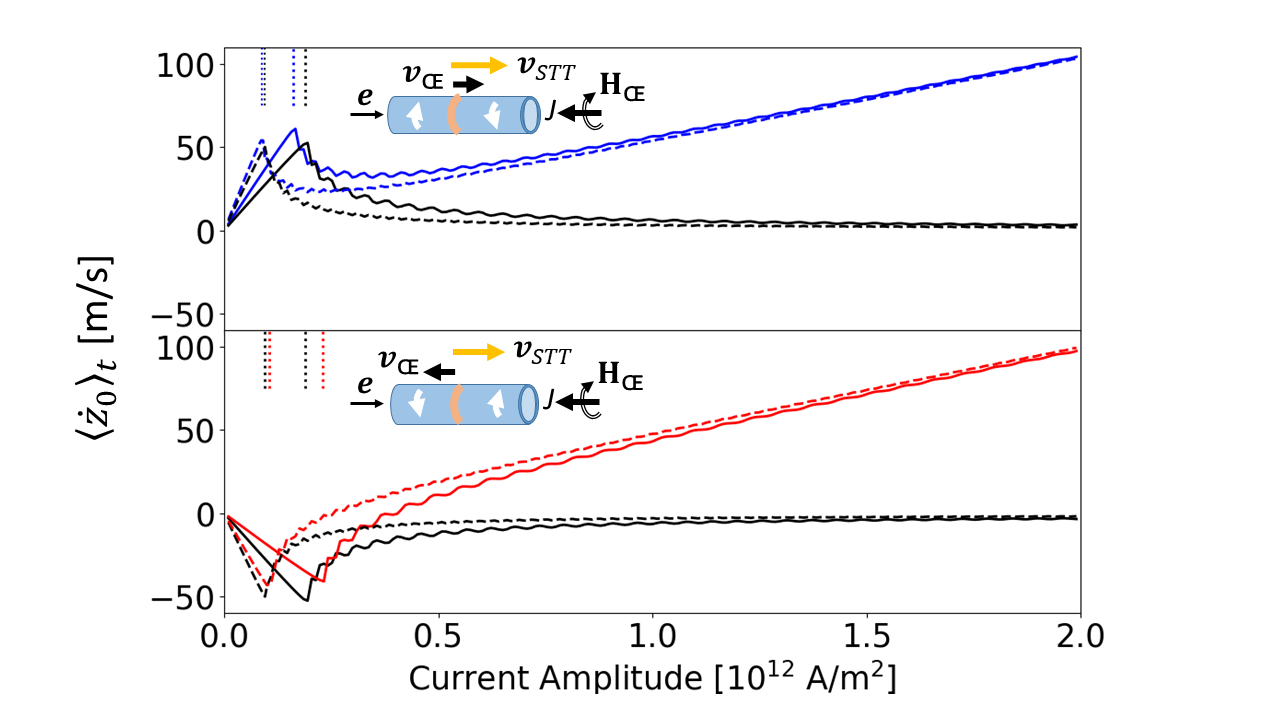}
  \caption{ Time-averaged DW speed obtained with the $1$D  model for a Bloch DW versus the density of electric current and for two nanotube thicknesses ($\Delta R=\SI{10}{\nano\meter}$ and $\Delta R=\SI{20}{\nano\meter}$) and for three different configurations: disconsidering spin-transfer torques~(black curves), with spin-transfer torque and \OErsted field cooperating~(red curve), and with spin-transfer torque and \OErsted competing~(blue curve). The vertical dotted lines highlight the critical current $\JW$ defined in Eq.\eqref{Eq: Critical Oersted field}. The following parameters have been used: $K = \SI{45E4}{\joule\per\cubic\meter}$, $\Rext = \SI{50}{\nano\meter}$ and $\alpha=0.02$.}
 \label{fig: Average_velocity_vs_alpha_1D_model_azimuthal_bloch_wall}
\end{figure}

To illustrate Bloch DW behavior  we plot in Fig.\ref{fig: Average_velocity_vs_alpha_1D_model_azimuthal_bloch_wall} the time-averaged domain DW speed as a function of the electric current amplitude, according to Eqs. \eqref{Eq:equation of motion collective variables model lambda}-\eqref{Eq:equation of motion collective variables model chi}.

In the steady-state regime below the critical current~$\JW$, the DW velocity varies linearly with the current and is largely determined by the effect of the \OErsted field, due to the demultiplicating $1/\alpha$ coefficient for the mobility~[Eq.\eqref{Eq:Velocity_domain_wall_small_Orsted_field_Bloch}]. Nevertheless, the mobility and thus the maximum speed reachable is much lower than for Néel DWs with low anisotropy, due to much narrower thickness, $\lex$, see Fig. \ref{fig:Energy_analytic_vs_smulations_ground_state}(c),(d). Also, similar to the case of Néel DWs, the direction of motion in steady regime is dictated by the sign of circulation of the azimuthal domains, so that two consecutive DWs in a nanotube are expected to move along opposite directions. The relative difference of slopes between curves with cooperating or competing \OErsted and spin-transfer-torque-induced contributions is more pronounced in comparison with the Néel DWs considered in~Fig.\ref{fig: Average_velocity_vs_alpha_1D_model_azimuthal_neel_wall}. This is related to the much narrower width of the Bloch DWs so that field-induced and spin-transfer-torque-induced driving forces contribute comparably.

Above $\JW$, the Bloch wall dynamics is largely dominated by the spin-transfer torque. It means that in contrast to \OErsted field dominated dynamics the direction of the DW does not depend on the initial orientation of the domains and follows the electron flow direction. This allows better control of DW manipulation if needed. The DW average speed is asymptotically proportional to $C_{\pm} U$ and, as expected, almost does not depend on the shell thickness. For parameters studied here $1$D model predicts the DW speed increase by $50$ m/s for an increase of the applied current of $\SI{1E12}{\ampere\per\meter\squared}$.

\begin{figure*}[t]
\centering
\includegraphics[scale=0.45]{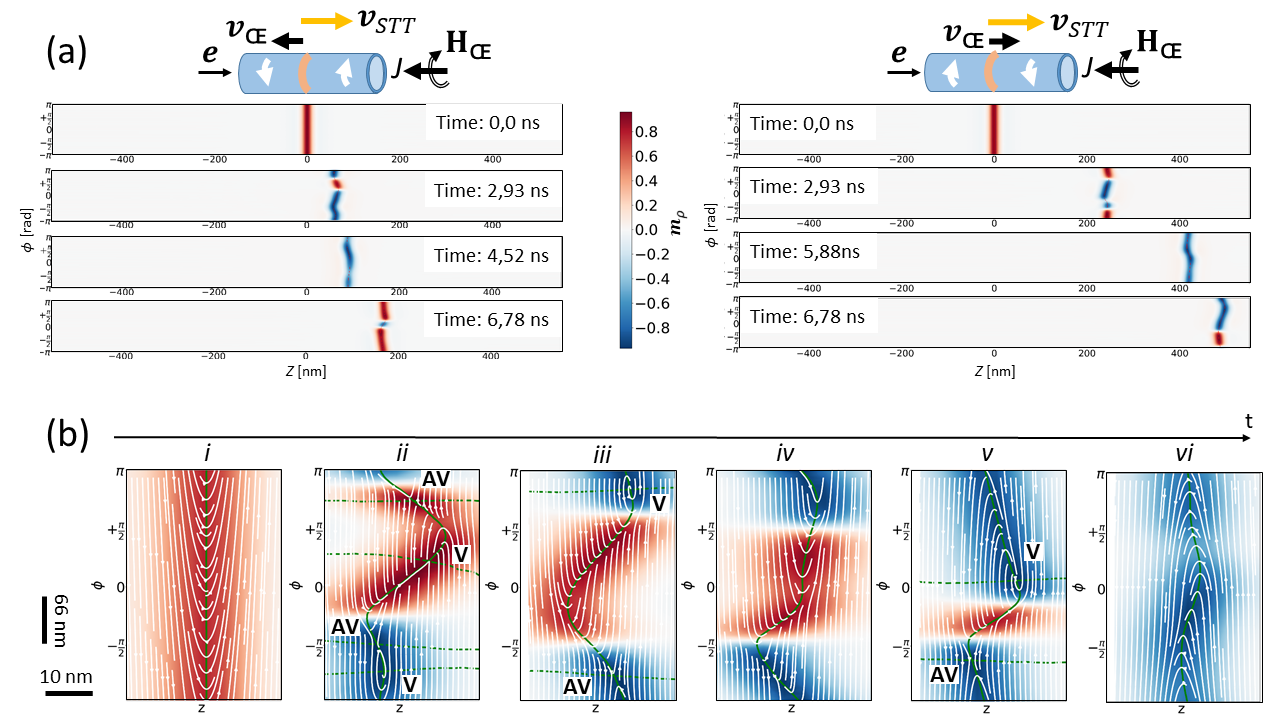}
  \caption{(a) Outer-surface maps for the micromagnetic simulation of the dynamics of a Bloch DW above the critical current. Two situations are illustrated: competing \OErsted and spin-transfer torque~(left panel) and cooperating \OErsted and spin-transfer torque~(right panel). The colors indicate the radial magnetization component $m_\rho$, and the two lateral scales are identical . The system is a Permalloy nanotube with $J = \SI{1E12}{\ampere\per\meter\squared}$, $K = \SI{25E4}{\joule\per\meter\squared}$, $\Rext = \SI{50}{\nano\meter}$, $\Delta R = \SI{10}{\nano\meter}$ and $\alpha=0.02$. (b)~Enlargement of the magnetic texture around the DW taken at different times during the wall polarity switching process and corresponding to the situation shown above on the left side. Note that the axial scale has been expanded, to allow for a better visibility. The white lines correspond to a stream plot displaying the in-plane magnetization vector field $(m_z,m_\phi)$ and the green lines indicate the iso-contours $m_\phi = 0$ (full lines) and $m_z=0$ (dashed lines). In addition, vortex (V) and anti-vortex (AV) are indicated, each standing at the intercept between green dashed and full lines, i.e. $m_\rho = \pm 1$.}
  \label{fig:Bloch_carthography}
\end{figure*}

The comparison between the $1$D analytic model and micromagnetics demonstrates a qualitatively similar behavior for parameters considered here: low critical currents values $J_{\textrm{w}}$~ [Fig. \ref{fig:shema_1D_model_walker_field}(b)], steady regime below $J_{\textrm{w}}$ and spin-transfer-torque-dominated Walker regime above $J_{\textrm{w}}$ (Fig. \ref{fig:Bloch_carthography}).
However, the estimation of the DW speed for the applied current of $\SI{1E12}{\ampere\per\meter\squared}$, for instance, gives some difference in comparison to $1$D model for \OErsted and spin-transfer torque competing and cooperating configuration: $30~\rm{m/s}$ and $64~\rm{m/s}$, respectively. In contrast to $1$D model, this indicates that the  contribution from the \OErsted field is approximately $3$ times lower than that of spin-transfer torque and not completely negligible. 

Simulations allow to grasp the complexity of the underlying micromagnetic processes above $\JW$, well beyond the $1$D model. In Fig. \ref{fig:Bloch_carthography}, we show the motion of a Bloch wall for \OErsted and spin-transfer torque competing and cooperating configurations. The direction of the Bloch wall movement is imposed by the electron flow. We observe the switch of the Bloch wall polarity with intermediate Bloch-Néel transformation during its movement as expected in the Walker regime. It is clear that the process of transformation from Bloch DW to Néel DW is not the coherent rotation of the DW core, directly mapped by 1D model. 

Although the DW speed itself depends on the cooperation or competition between  \OErsted field and spin-transfer torque~[see the timeframe in Fig.\ref{fig:Bloch_carthography}(a)], the common feature for both configurations is the nucleation of a vortex-antivortex pairs during Bloch wall polarity switching. As shown in Fig.\ref{fig:Bloch_carthography}(b), initially the Bloch DW has a positive polarity ($m_\rho>0$). Magnetization is progressively tilting toward the axial direction until the creation of one or more pairs of vortex-antivortex encompassing a small locus with $m_\rho<0$. These move along opposite directions, contributing to progressively reversing the average DW polarity, until their mutual annihilation. 

Similar but not identical qualitative behavior have been described in other simulations, such as in strips\citep{bib-NAK2003}, thin films\cite{bib-KRI2019} or a vortex wall between head-to-head domains in nanotubes\citep{bib-YAN2012,bib-HER2016}. The nucleation/annihilation processes can be formalized on the topological basis of the so-called skyrmion number preservation\citep{bib-TRE2007}.

\section{Conclusion}

We predict features of the statics and current-induced motion of domain walls in ferromagnetic nanotubes with azimuthal magnetization. Experimental reports are emerging on such tubes and walls, however not covered by existing theories, considering axial magnetization as expected for a soft magnetic material. We combine theory based on an analytical 1D model to draw trends and highlights the physics at play, with micromagnetic simulations for an accurate description. While thin-shell tubes may be mapped to a flat strip by a gedanken unrolling, the resulting situation would be analogous to domains with magnetization transverse to the strip, which has not been covered yet. Besides, new physics arises specifically in tube versus strips, case due to curvature-induced exchange-related anisotropy, volume magnetostatic energy breaking the inversion symmetry, and most importantly the interplay of an \OErsted field with spin-transfer torque. We discuss step by step which features are analogous to strips, and which are specific to tubes.

Key to all results is the transition from Néel to Bloch as ground states for domain walls, the former favored for large radius, low azimuthal microscopic anisotropy and low shell thickness. This duality has a key impact on the dynamics as the Walker current is related to the restoring force between the two types of walls, vanishing at the transition due to a soft mode. Below the Walker current the wall motion is mostly driven by the \OErsted field, so that successive domain walls move along opposite directions. In this regime, Néel walls have a large width especially in the experimentally-relevant range of moderate azimuthal anisotropy, so that they may reach a speed close to $\SI{1}{\kilo\meter\per\second}$, while Bloch walls move one order of magnitude slower. Above the Walker current, the motion is mostly driven by spin-transfer torque, with similar speeds in both cases and all domain walls moving along the electron flow. Therefore, a peculiar outcome is the change of direction of motion of domain walls across the Walker current, when the \OErsted field and spin-transfer torques compete.

\section{Acknowledgements}

We acknowledge useful discussions with Michael Schöbitz~(SPINTEC, Grenoble, France) and financiel support from ANR projects ANR-17-CE24-0017  and ANR-JCJC-MATEMAC-3D.

\bibliographystyle{apsrev4-2}

\bibliography{Fruche8}

\appendix

\section{Energy functional minimization}
\label{Appendix:Euler_equation}
In our $1$D model, the functions $\theta(z)$ and  $\Psi(z)$ that minimize the energy functional \eqref{Eq:Energy functional 1D model for ground state} are the solutions of the following Euler-Lagrange equations:
\begin{align}
\frac{\partial \mathcal{E}}{\partial \theta} - \frac{\partial}{\partial z} \frac{\partial \mathcal{E}}{\partial \theta'} = 0, ~~~~~~ \frac{\partial \mathcal{E}}{\partial \Psi} - \frac{\partial}{\partial z} \frac{\partial \mathcal{E}}{\partial \Psi'} = 0\;,
\label{Eq:Euler-Lagrange}
\end{align}
with partial derivatives $\theta' = \partial \theta/\partial z$, $\Psi' = \partial \Psi/\partial z$, energy functional density $\mathcal{E}=\mathcal{E}[\bm{m}]$ and $E[\bm{m}]=\int \partial V \mathcal{E}[\bm{m}]$.

\section{Magnetization profile}
\label{Appendix:1D_model_profile}
By setting $\Psi = \pi/2$, the differential equations \eqref{Eq:set of differential equation theta} - \eqref{Eq:set of differential equation psi} reduce to:
\begin{align}
\frac{\partial^2\theta}{\partial z^2} & =  -\frac{\sin  \theta \cos \theta}{\Lambda_0^2}
\label{Appendix_Eq:Differential_equation_theta}.
\end{align}
By multiplying Eq. \eqref{Appendix_Eq:Differential_equation_theta} by $\partial \theta/\partial z$ and by integrating the differential equation over z, one obtains:
\begin{align}
\left(\theta '\right)^2 - \frac{\cos^2(\theta)}{\Lambda_0^2} = \mathrm{Cste}.
\label{Appendix_Eq:Differential_equation_theta_simplified}
\end{align}
For $\mathrm{Cste} =0$ the solution of Eq. \eqref{Appendix_Eq:Differential_equation_theta_simplified} reads

\begin{align}
\theta_{\pm}\left(z\right) = 2\tan^{-1}\left[\tanh \left( \pm \frac{z}{2 \Lambda_0} \right) \right],
\end{align}
and the domain DW is located at $z=0$. The $\theta_+$ solution corresponds to the N\'{e}el DW between two azimuthal domains, for which $m_\phi$=$\pm 1$ at $z$=$\pm\infty$, and $\theta_-$ to the N\'{e}el DW with $m_\phi$=$\mp$ at $z$=$\pm\infty$.

Similar, by setting $\theta = \pi/2$, the solution corresponding to the Bloch DW reads:
\begin{align}
\Psi_{\pm}\left(z\right) = 2\tan^{-1}\left[\tanh \left( \pm \frac{z}{2 \lex} \right) \right].
\end{align}
\\

\section{Equation of motion}
\label{Appendix:Equation_of_motion}
To derive the time evolution of the collective variables \eqref{Eq:equation of motion collective variables model lambda}-\eqref{Eq:equation of motion collective variables model chi}, we followed the approach used in ref.\citenum{bib-LAN2010}. We started with  the LLG equation \eqref{Eq:LLG} in spherical coordinates
\begin{align}
\left \{
\begin{array}{r c l}
      \dot{\mathcal{Q}} &=& \displaystyle  \frac{\gamma_0}{\mu_0 \Ms} \frac{\delta \mathcal{E}}{\delta \mathcal{P}} +  \left( 1- \mathcal{Q}^2 \right) \left[ \alpha \dot{P} + \beta U  \frac{\partial \mathcal{P}}{\partial z} \right]   - U \frac{\partial \mathcal{Q}}{\partial z} \;, \\ \\
          \dot{P} &=& \displaystyle  -\frac{\gamma_0}{\mu_0 \Ms} \frac{\delta \mathcal{E}}{\delta \mathcal{Q}} -  \frac{1}{1- \mathcal{Q}^2 } \left[ \alpha \dot{\mathcal{Q}} +  \beta U \frac{\partial \mathcal{Q}}{\partial z} \right] - U \frac{\partial \mathcal{P}}{\partial z}\;,
\end{array}
\right .
\label{Appendix_Eq:Eq LLG Q,P}
\end{align}
where we defined $\mathcal{Q} =  \cos \theta$ and $\mathcal{P} = \Psi$.
Equation \eqref{Appendix_Eq:Eq LLG Q,P} has a Hamiltonian structure, \ie, $\dot{\mathcal{P}} = - \frac{\partial\mathcal{H}\left(\mathcal{Q},\mathcal{P}\right)}{\partial \mathcal{Q}}$,  $\dot{\mathcal{Q}} =  \frac{\partial\mathcal{H}\left(\mathcal{Q},\mathcal{P}\right)}{\partial \mathcal{P}}$ with the Hamiltonian density of the system
$\mathcal{H} = \frac{\gamma_0}{ \mu_0 \Ms} \mathcal{E} \left[\mathcal{P}, \mathcal{Q} \right]$ and its Lagrangian density:

\begin{align}
\mathcal{L} = \mathcal{P} \dot{\mathcal{Q}} - \frac{\gamma_0}{\mu_0\Ms}\mathcal{E} \left[\mathcal{P}, \mathcal{Q} \right].
\label{Appendix_Eq:Eq Lagrangian density LLG}
\end{align}

The equations of motion of the system are given by:
\begin{align}
\frac{\delta S }{ \delta \mathcal{Q}} = \frac{\delta R }{ \delta \dot{\mathcal{Q}}} ~~~\textrm{and}~~~ \frac{\delta S }{ \delta \mathcal{P}} = \frac{\delta R}{ \delta \dot{\mathcal{P}}},
\label{Appendix_Eq:Eq_of motion LLG from action}
\end{align}
where $S = \int \partial t L$ is the action of the system and $L = \int \partial\bm{r} \mathcal{L}$ is the Lagrangian.
The dissipation function of the system $R = \int  \partial t ~\partial\bm{r} \mathcal{R}$ reads
\begin{align}
\mathcal{R}  &= \frac{\alpha}{2} \left[ \frac{\dot{\mathcal{Q}}^2}{1- \mathcal{Q}^2} + \left(1- \mathcal{Q}^2\right)\dot{\mathcal{P}}^2  \right]
+
U  \left[ \frac{\partial \mathcal{P}}{\partial z} \dot{Q}
-
  \frac{\partial Q}{\partial z} \dot{\mathcal{P}} \right] \nonumber \\
&+ \beta U
  \left[  \frac{\dot{\mathcal{Q}} \frac{\partial Q}{\partial z}}{1- \mathcal{Q}^2} +  \left(1- \mathcal{Q}^2\right)\dot{\mathcal{P}} \frac{\partial \mathcal{P}}{\partial z} \right].
\label{Appendix_Eq:Eq dissipation LLG}
\end{align}
The next step consists in finding a suitable Ansatz for the spatial profile of magnetization during the dynamics, which contains a given number of time-dependent collective variables $ \chi_{i}(t)$, \ie,  $ \mathcal{Q} \left( \bm{r},t\right) = \mathcal{Q}\left[\bm{r}, \chi_{i}(t) \right]$ and $\mathcal{P} \left( \bm{r},t\right) = \mathcal{P}\left[ \bm{r}, \chi_{i}(t) \right]$. Then the set of equations \eqref{Appendix_Eq:Eq_of motion LLG from action} is replaced by a new set of differential equations:
\begin{align}
 \frac{\partial L}{\partial \chi_{i}} - \frac{\partial}{\partial t} \frac{\partial L}{\partial \dot{\chi_{i}}} = \frac{\partial R}{\partial \dot{\chi_{i}}} - \frac{\partial}{\partial t} \frac{\partial R}{\partial \ddot{\chi_{i}}},
\label{Appendix:Eq of motion LLG from action with collective variables}
\end{align}
one for each collective variable. To benefit from this method it is essential to find an analytic an accurate expression for the Lagrangian and the dissipation function. The latter are obtained by integrating over space the Lagrangian density and the dissipation density function. \\

Using the Ansatz \eqref{Eq:Anzatz_dynamics_azimuthal_neel_wall},
we find the following expression for $\mathcal{Q}$ and $\mathcal{P}$:
\begin{align}
\mathcal{Q}   = \frac{\cos \chi(t)} {\cosh\left[ \frac{z - z_0(t)}{\Lambda (t)}\right]},~~ \mathcal{P}= \tan^{-1} \left\{ \frac{\sinh\left[ \frac{z - z_0(t)}{\Lambda (t)}\right] }{\sin \chi(t)}\right\}.
\label{Appendix:Eq Q P first ansatz  azimuthal neel wall}
\end{align}
The Lagrangian of the infinitely long-system associated to this Ansatz reads
\begin{align}
L\left(z_0,\Lambda,\chi \right)=& - \frac{2\gamma_0AS\Lambda}{\mu_0\Ms}  \left(
\frac{1}{\Lambda^2} - \frac{\cos^2 \chi}{\lambda^2} + \frac{\sin^2\chi}{\lex^2}\right. \nonumber \\
& \left. + \frac{\cos^2\chi}{W^2} +  \frac{2z_0}{\Lambda \lOE^2}
\right)
 + 2S   \dot{z_0} \tan^{-1}\left(\frac{1}{\tan\chi}\right)\;
 \label{Appendix:Eq_Lagrangian}
\end{align}
and the dissipation function:

\begin{align}
R\left(z_0,\Lambda,\chi \right)
 &=
  \alpha S \left( \frac{ \dot{z_0}^2}{\Lambda}   +   \Lambda\dot{\chi}^2  +\frac{\pi^2\dot{\Lambda}^2}{12 \Lambda}  \right) \nonumber \\
 &  +  2SU \dot{\psi} + 2SU\beta \left( \frac{z_0}{\Lambda^2} \dot{\Lambda} - \frac{1}{\Lambda} \dot{z_0}\right).
\label{Appendix:Eq_Dissipation_function}
\end{align}
Note that some constant terms have been skipped since they do not contribute to the equations of motion. Finally, using equations \eqref{Appendix:Eq of motion LLG from action with collective variables} together with the expressions \eqref{Appendix:Eq_Lagrangian} and \eqref{Appendix:Eq_Dissipation_function}, we find the equation of motion \eqref{Eq:equation of motion collective variables model lambda}-\eqref{Eq:equation of motion collective variables model chi} of the main text.

\end{document}